\documentclass[12pt]{article}
\usepackage{latexsym}
\usepackage{amsmath,,calrsfs}
\usepackage{amsfonts}
\usepackage{amssymb}
\usepackage{amscd}
\usepackage{fancybox}
\usepackage{cite}
\usepackage{amsmath,amsfonts,amsbsy}
\usepackage{pstricks,pst-node}
\usepackage[small,bf,hang]{caption2}
\usepackage{psfrag}
\usepackage{graphicx}

\usepackage{float}

\psset{unit=1.3cm,linewidth=.5pt,radius=.2}  

\usepackage{enumitem}                       
\usepackage{multirow}                     
\usepackage{float}                          
\usepackage{lscape}                         
\usepackage{bm}

\def\hybrid{\topmargin -20pt    \oddsidemargin 0pt
        \headheight 0pt \headsep 0pt
        \textwidth 6.25in       
        \textheight 9.5in       
        \marginparwidth .875in
        \parskip 5pt plus 1pt   \jot = 1.5ex}

\hybrid

\newcommand{\beq}{\begin{equation}}
\newcommand{\eeq}{\end{equation}}
\newcommand{\bi}{\begin{itemize}}
\newcommand{\ei}{\end{itemize}}
\newcommand{\bt}{\begin{tabular}}
\newcommand{\et}{\end{tabular}}
\newcommand{\bc}{\begin{center}}
\newcommand{\ec}{\end{center}}

\def\theequation{\arabic{section}.\arabic{equation}}

\newcommand{\ft}[2]{{\textstyle {\frac{#1}{#2}} }}

\newcommand{\be}{\begin{equation}}
\newcommand{\ee}{\end{equation}}
\newcommand{\bea}{\begin{eqnarray}}
\newcommand{\eea}{\end{eqnarray}}
\newcommand{\ba}{\begin{array}}
\newcommand{\ea}{\end{array}}

\def\bbox{{\,\lower0.9pt\vbox{\hrule \hbox{\vrule height 0.2 cm
\hskip 0.2 cm \vrule height 0.2 cm}\hrule}\,}}
\newcommand{\dsl}{\pa \kern-0.5em /}

\begin{document}

\begin{titlepage}
\begin{center}

\hfill UG-09-06 \\
\hfill DAMTP-2009-36

\vskip 2cm

{\Large \bf  More on Massive 3D Gravity\\}

\vskip 2cm

{\bf Eric A.~Bergshoeff\,$^1$, Olaf Hohm\,$^1$ and Paul K.~Townsend\,$^2$} \\

\vskip 30pt

{\em $^1$ \hskip -.1truecm Centre for Theoretical Physics, University of Groningen, \\
Nijenborgh 4, 9747 AG Groningen, The Netherlands \vskip 5pt }

{email: {\tt E.A.Bergshoeff@rug.nl, O.Hohm@rug.nl}} \\

\vskip 15pt

{\em $^2$ \hskip -.1truecm Department of Applied Mathematics and
Theoretical Physics,\\
Centre for Mathematical Sciences, University of Cambridge,\\
Wilberforce Road, Cambridge, CB3 0WA, U.K. \vskip 5pt }

{email: {\tt P.K.Townsend@damtp.cam.ac.uk}} \\

\end{center}

\vskip 1cm

\begin{center} {\bf ABSTRACT}\\[3ex]
\end{center}

We explore the space of static solutions of  the recently discovered
three-dimensional `New Massive Gravity' (NMG),  allowing for either
sign of the Einstein-Hilbert term and a cosmological term
parametrized by a dimensionless constant $\lambda$.  For
$\lambda=-1$ we find  black hole solutions asymptotic (but not
isometric) to the unique (anti) de Sitter  vacuum, including
extremal black holes that interpolate between this vacuum and
(a)dS$_2 \times S^1$.  We also investigate  unitarity of  linearized
NMG in (a)dS vacua. We find unitary theories for some dS vacua,  but
(bulk) unitarity in adS implies  negative central charge of the dual
CFT,  except for $\lambda=3$ where the central charge vanishes and
the bulk gravitons are replaced by `massive photons'.  A similar
phenomenon is found in the massless limit of NMG, for which the
linearized equations become equivalent to Maxwell's equations.

\begin{minipage}{13cm}
\small

\end{minipage}




\vfill

\end{titlepage}

\section{Introduction}\setcounter{equation}{0}

We recently found a novel three-dimensional ($3D$) gravity model that propagates massive positive-energy spin 2 modes, of both helicities $\pm2$,  in a Minkowski vacuum \cite{Bergshoeff:2009hq};  in other words, an interacting, and generally covariant, extension of the  Pauli-Fierz (PF) theory for massive spin $2$ in $3D$.  The action for this  `new massive gravity' (NMG) is the sum of a `wrong-sign'  Einstein-Hilbert (EH) term and a particular higher-derivative term, which introduces a mass parameter $m$.  Models of this type are known to be renormalizable in four dimensions \cite{Stelle:1976gc}, and this implies power-counting super-renormalizability in three dimensions. Unitarity has since been confirmed \cite{Nakasone:2009bn,Deser:2009hb}, as has
super-renormalizability  \cite{Oda:2009ys}, so that NMG must now be considered a promising candidate for
a fully consistent theory of quantum gravity, albeit in three dimensions and with massive gravitons.

We also found a parity-violating extension of NMG to a `general
massive gravity' (GMG) in which the $\pm2$ helicity modes propagate
with different masses $m_{\pm}$; NMG is recovered by setting
$m_+=m_-=m$ while the limit $m_-\to\infty$ yields the  well-known
`topologically massive gravity' (TMG) \cite{Deser:1981wh}, which
propagates only a single mode of helicity $2$.   Also considered
briefly in \cite{Bergshoeff:2009hq} was the extension of  GMG to
allow for a cosmological constant; this can be chosen to be
proportional to $\lambda m^2$, where $\lambda$ is a  new {\it
dimensionless} parameter.  Studies of `cosmological TMG' (CTMG)
\cite{Deser:2002iw,Li:2008dq,Carlip:2008jk} suggest that one should
also allow for either  sign of the EH term in `cosmological' GMG
(CGMG), so it is convenient to introduce a sign $\sigma$ such that
$\sigma=1$ for the `right-sign' EH term and $\sigma=-1$ for the
`wrong-sign' EH term\footnote{We use the terms `right-sign' and
`wrong-sign' because the actual signs are convention dependent.}.
For the `mostly plus' signature convention that we will use here,
the CGMG action  is \be\label{ouraction}
 S_{CGMG}[g]= \frac{1}{\kappa^{2}}\int \! d^3 x\, \left\{ \sqrt{|g|} \left[ \sigma R  +  \frac{1}{m^{2}} K
-2\lambda m^2\right]  +\frac{1}{\mu}{\cal L}_{\rm LCS}\right\}\, ,
 \ee
 where
 \be\label{defK}
|g| = -\det g \, , \qquad K =  R_{\mu\nu}R^{\mu\nu} - \frac{3}{8}R^2\, ,
\ee
and the Lorentz-Chern-Simons (LCS) term is
 \be
  {\cal L}_{\rm LCS}=\frac{1}{2}\varepsilon^{\mu\nu\rho} \left[
  \Gamma^\alpha_{\mu\beta} \partial_\nu \Gamma^\beta_{\rho\alpha} +
  \frac{2}{3} \Gamma^\alpha_{\mu\gamma}
  \Gamma^\gamma_{\nu\beta}\Gamma^\beta_{\rho\alpha} \right] \, ,
 \ee
where $\Gamma$ is the usual Levi-Civita connection for the spacetime metric $g$.  The Riemann curvature tensor is determined, in $3D$, by the Ricci tensor, which is
 \be
  R_{\mu\nu} \ \equiv \ R_{\rho\mu}{}^{\rho}{}_{\nu} \ = \
  \partial_{\rho}\Gamma^{\rho}_{\mu\nu}-\partial_{\mu}\Gamma^{\rho}_{\rho\nu}
  +\Gamma_{\rho\lambda}^{\rho}\Gamma_{\mu\nu}^{\lambda}-\Gamma_{\mu\lambda}^{\rho}\Gamma_{\rho\nu}^{\lambda}\, .
 \ee
It would be possible to remove the higher-derivative  term $K$ by a local redefinition of the metric, but this would just introduce an infinite series of yet higher-derivative terms\footnote{Linearization would then  yield, for $\sigma=-1$ and $\lambda=0$, another higher-derivative model equivalent to the PF theory, so this redefinition is perhaps not without  interest.}. To  remove {\it all} higher-derivative terms (other than the LCS term) would require an inadmissible non-local redefinition.

In this paper we focus on the CNMG model obtained as the
$\mu\to\infty$ limit of the CGMG model, although all static
solutions of CNMG  are also solutions of the CGMG  for any value of
$\mu$ because the LCS term plays no role for static solutions. We
shall make use of the following  variant of the CNMG action,
involving an auxiliary symmetric tensor field $f$
\cite{Bergshoeff:2009hq}: \be\label{CNMG}
 S_{CNMG}[g,f]= \frac{1}{\kappa^{2}}\int \! d^3 x\, \sqrt{|g|} \left[ -2\lambda m^2 + \sigma R  + f^{\mu\nu}G_{\mu\nu}
 - \frac{1}{4}m^2 \left(f^{\mu\nu}f_{\mu\nu} -f^2\right) \right] \, .
 \ee
The $\mu\to\infty$ limit of (\ref{ouraction}) is recovered after  elimination of $f$ by its field equation
 \bea\label{fsol}
  f_{\mu\nu} \ = \ \frac{2}{m^2} S_{\mu\nu}\, , \qquad S_{\mu\nu} \equiv
 R_{\mu\nu}-\frac{1}{4}R g_{\mu\nu}\, .
 \eea
 The tensor $S$ appears in the conformal approach to supergravity as the gauge field corresponding to conformal boosts \cite{Kaku:1978nz,Uematsu:1984zy}, and has been called the `Schouten' tensor  \cite{Gover:2008sw}.

As an aside, we remark here that  the action (\ref{CNMG}) has the feature that it allows one to take the $m\to 0$ limit by defining $\tilde f_{\mu\nu} = m^2 f_{\mu\nu}$ and keeping $m\kappa \equiv \beta$ fixed.  This
yields  the action
 \be\label{masslessact}
 S_{massless}[g,\tilde{f}] = \frac{1}{\beta^2}\int \! d^3 x\, \sqrt{|g|}  \left[ \tilde f^{\mu\nu} G_{\mu\nu} - \frac{1}{4} \left(\tilde f^{\mu\nu} \tilde f_{\mu\nu} - \tilde f^2\right)\right] \, ,
 \ee
and elimination of $\tilde f$ now yields \be\label{pureK} S[g] =
\frac{1}{\beta^2} \int \! d^3 x \, \sqrt{|g|}\, K\, .
\ee
The quadratic approximation to this `pure' higher-derivative model was
recently studied by Deser \cite{Deser:2009hb},  who shows  that
there is a single, physical, massless propagating mode.  We confirm this result
here by showing that the linearized action in this case is equivalent to a Maxwell action
for a vector field, which is itself on-shell equivalent to the action for a massless scalar.
The `Weyl-invariance' discussed in \cite{Deser:2009hb} is realized as a
Maxwell gauge invariance in this context. However, this gauge invariance is broken by
the interactions, so the consistency of the non-linear `pure-K' model is problematic. We investigate this issue
here (in an appendix) because it turns out to be relevant to the $\lambda=3$ case of the CNMG model. We also investigate
the `cosmological' extension of the `pure-K'  model obtained by the addition of a cosmological term

By definition, any
maximally-symmetric vacuum of CGMG is such that \be\label{maxsym}
G_{\mu\nu} = -\Lambda g_{\mu\nu} \, , \ee where $G$ is the Einstein
tensor, and  $\Lambda$ is a constant (the sign is chosen such that
$\Lambda>0$ for de Sitter (dS) spacetime and $\Lambda<0$ for anti-de
Sitter (adS) spacetime).  It was shown in \cite{Bergshoeff:2009hq}
that  such configurations solve the CGMG field equations if, and
only if, $\Lambda$ solves the quadratic equation
 \be\label{Lambdarel}
 \Lambda^2+4m^2\sigma\Lambda -4\lambda m^4 \ = \ 0\, .
 \ee
In other words, using $\sigma^2=1$,
 \be
 \Lambda= -2m^2 \left[\sigma \pm \sqrt{1+\lambda}\right]\, .
 \ee
We see that there is a Minkowski vacuum when $\lambda=0$, but also a
dS ($\sigma=-1$) or adS ($\sigma=1$) vacuum\footnote{That
higher-derivative gravity theories generically permit cosmological
solutions even without an explicit cosmological constant in the
action is well-known \cite{Boulware:1985wk}.}. For $\lambda>-1$,
there are two solutions  and hence two possible maximally symmetric
vacua. These two vacua merge at $\lambda=-1$ to give a unique vacuum
that is adS if $\sigma=1$ and dS if $\sigma=-1$. For $\lambda<-1$
there is no maximally-symmetric solution.

Maximally-symmetric vacua constitute a special subclass of the more general class of static solutions, which we investigate here. Any static metric can be put in the form
\be\label{static}
ds^2 = -ab^2 dt^2 +  a^{-1} dr^2 + \rho^2 d\theta^2\, ,
\ee
for some functions $(a,b,\rho)$ of the one independent variable $r$.
The special case in which $b^2= \rho^2/a$  yields the `domain-wall'  metrics
\be\label{dwmetric}
ds^2 = \rho^2\left(-dt^2 + d\theta^2\right) + a^{-1} dr^2\, .
 \ee
These have an additional $2D$ `worldvolume' Poincar\'e isometry. We shall show that all static solutions of this special type are locally equivalent to an adS vacuum. This is perhaps disappointing but one interesting feature
emerges from this analysis: for $\lambda=3$ the domain-wall `energy' (the negative of  the Lagrangian) is
a perfect square so that the adS vacuum for this case saturates a Bogomolon'yi-type bound,  suggesting
a possible connection with supersymmetry.

The general static metric includes static black hole spacetimes,  and whenever there exists an  adS vacuum we know that we must also find BTZ black holes \cite{Banados:1992wn} because these are locally isometric to the adS vacuum.  An interesting question is whether there exist static black holes that are {\it not} locally isometric to the (a)dS vacuum.
There are two aspects to this problem: first one must find static solutions that are not locally (a)dS, and then one must
determine which of these is non-singular on and outside an event horizon.  We reduce the first part of this problem
to the solution of a pair of coupled ODEs for the functions $(a,b)$ and we find the general solution of these equations  for which $b\equiv 1$. The only black hole solutions that we find in this way for generic $\lambda$ are the BTZ black holes but we find a new class of black holes  for  $\lambda=-1$. For $\sigma=-1$ they are analogous to dS black holes in that there is both a black hole horizon and a cosmological horizon. For $\sigma=1$ they are asymptotically adS black holes, generically with non-zero surface gravity  but there is an `extremal' limit that yields a black hole with zero surface gravity (and hence zero Hawking temperature). In this case the black hole interpolates between the adS$_3$ vacuum at infinity and, near the horizon,  an adS$_2 \times S^1$ Kaluza-Klein (KK) vacuum  found previously by Clement \cite{Clement:2009gq}.
One can also take an extremal limit for  $\sigma=-1$; in this case the black hole and cosmological singularities coincide. In appropriate new coordinates, the result is a
time-dependent  cosmological solution that is asymptotic to a dS$_2\times S^1$ KK vacuum in the far past.  We verify directly that these KK vacua are indeed solutions of the CGMG equations at $\lambda=-1$.  An interesting corollary is that the dimensional reduction of the $\lambda=-1$ CGMG theory yields a 2D gravitational theory with either a dS$_2$ vacuum (if $\sigma=-1$) or an  adS$_2$ vacuum (if $\sigma=1$).

We should mention here  that  stationary  non-BTZ  black hole
solutions of CNMG have been found previously
\cite{Clement:2009gq,Clement:2009ka}, but these do not include any
new {\it static} black hole solutions.  A class of `adS wave'
solutions has been found in \cite{AyonBeato:2009yq}.  While the
original version of this paper was in the stages of completion, we
were informed  by R.~Troncoso  that he and collaborators had
independently found new solutions at the special point $\lambda=-1$,
and their paper has now appeared \cite{Oliva:2009ip}.  We have also
been informed by A.~Maloney that he and A. Wissanji have found new
stationary black hole solutions for arbitrary $\lambda$.

Another purpose of this paper is to examine the stability of the
(a)dS vacua of CNMG, and to determine which of the linearized field
theories are unitary. Here we recall one of the main results of
\cite{Bergshoeff:2009hq}: for $\sigma=-1$ and $\lambda=0$,  the
linearization about the Minkowski vacuum yields a unitary field
theory propagating two massive, and non-tachyonic, modes of spin 2,
with helicities $2$ and $-2$. The question we address is whether
these nice features persist in other maximally-symmetric vacua.

For adS backgrounds we will also discuss aspects of the dual
conformal field theory (CFT) on the boundary;  in particular we determine the regions in
parameter space where the central charges of the asymptotic Virasoro
algebra are positive, leading to unitary CFT's, and we will compare
with our findings for the bulk modes.

\section{The reduced energy functional}\setcounter{equation}{0}

Our main tool for finding static solutions is the principle of symmetric criticality \cite{PSC}; the application to GR has been discussed in \cite{Fels:2001rv}  and \cite{Deser:2003up}. This principle sanctions (under conditions that are satisfied in our examples) the substitution of an ansatz into the action provided that the ansatz is the most general one (up to diffeomorphisms) permitted by some group of isometries. We shall use this principle to obtain a reduced action for the functions $(a,b,\rho)$ that appear in the general static metric ansatz  (\ref{static}). A special feature of static metrics is that the LCS term does not contribute to the reduced action, so we can work with the simpler cosmological NMG model without losing generality. This simplification fails for the more general class of stationary metrics, so an investigation of stationary solutions using this method will be considerably more complicated.

The reduced action may be computed easily by using the fact that the Einstein tensor for the metric (\ref{static}) is
\be
G = - \frac{ab^2}{2\rho} \left[ \rho' a' +
2a\rho'{}'\right] dt^2 + \frac{\rho' \left(ba'+2ab'\right)}{2ab\rho}
dr^2 + \frac{\rho^2\left(ba' + 2ab'\right)'}{2b} d\theta^2\, ,
\ee
where the prime indicates differentiation with respect to $r$. Let us record here that this takes the following form for
the subclass of domain wall metrics with $b^2= \rho^2/a$:
\be
G = \rho\sqrt{a}\left(\rho\sqrt{a}\right)' \left[-dt^2 + d\theta^2\right] + \left(\frac{\rho'}{\rho}\right)^2 dr^2\, .
\ee
We shall work with the action (\ref{CNMG}), so we need to extend the static ansatz  (\ref{static}) to the auxiliary tensor field; we do this by taking it  to be time independent,  without further restriction, but its equation of motion
(\ref{fsol}) implies that it is also diagonal.  It is then convenient to also eliminate  from the action the diagonal components  $f^{rr}$ and $f^{tt}$; one is left with $f^{\theta\theta}$, which one can trade for the new function
 \be
c(r) = m^2 r^3 f^{\theta\theta}(r)\, .
\ee
Proceeding in this way,  and dropping an overall negative factor, we arrive at the `energy'  functional
\begin{eqnarray}\label{redfull}
E[a,b,c,\rho]  &=& \int \! dr \bigg\{ 2\lambda m^2 b\rho  - \sigma a\left( \rho' b'  - \rho'{}' b\right) + \nonumber \\
&&\left.  \frac{1}{2m^2\rho}\left[ \left(ba'+ 2ab'\right) \left(\left[
a\left(\rho'\right)^2\right]' + \rho c'\right) + bc^2 +
2\left(ab\rho'\right)' c \right]\, \right\}\, .
\end{eqnarray}
This  is invariant under the following gauge transformations:
\be\label{gaugetrans}
\delta_\xi \rho = \xi \rho' \, , \qquad \delta_\xi b = \left(\xi b\right)' \, , \qquad
\delta_\xi a = \xi a' -2 \xi' a\, , \qquad \delta c = \xi c'\, ,
\ee
from which we derive the identity
\be\label{ward}
\rho' \frac{\delta E}{\delta\rho} + c' \frac{\delta E}{\delta c} + 3a' \frac{\delta E}{\delta a}  + 2a \left(\frac{\delta E}{\delta a} \right)' - b \left(\frac{\delta E}{\delta b} \right)' \equiv 0\, .
\ee

\subsection{Domain walls}

One can further `reduce' the energy functional by the substitution
\be\label{bsub}
b^2= \rho^2/a\, ,
\ee
because this yields the generic `domain-wall'  metrics of (\ref{dwmetric}) compatible with  `worldvolume' boost invariance. After eliminating the variable $c$,  and discarding a boundary term, we find the new energy functional
\be\label{dwenergy}
E[a,\rho] = \int \! dr\,  \frac{1}{\sqrt{a}}\left\{ \frac{1}{6} \left[\frac{a\left(\rho'\right)^2 -6m^2\sigma}{m\rho}\right]^2 + 2\left(\lambda-3\right) m^2\rho^2 \right\}\, ,
\ee
which is manifestly positive for $\lambda>3$. For $\lambda=3$ the integrand of $E$ is a perfect square, which is evidently minimized when $\sigma=1$ by solutions of the first-order equation
\be
\rho' = \pm  \frac{\sqrt{6}\, m}{\sqrt{a}}\, .
\ee
This yields the adS vacuum of the model, and the way that we have found it suggests that it might be a
supersymmetric vacuum in the context of the supersymmetric extension of CGMG.

A remarkable feature of (\ref{dwenergy}) is that the equation of motion for $a$ is purely algebraic. Variation with respect to $a$ yields a quadratic equation for $\rho'$ with the solution
\be\label{aconstraint}
a \left(\rho'\right)^2 = \alpha^2 \rho^2 \, , \qquad
\alpha^2 = 2m^2 \left[\sigma \pm \sqrt{1+\lambda}\right] \, .
\ee
Because of gauge invariance, this implies the equation of motion for $\rho$ unless $\rho'=0$, but $\rho'=0$  is possible only for $\lambda=0$ and it leads only to the Minkowski vacuum (or identifications of it). Excluding this trivial case, (\ref{aconstraint}) is the only equation that we need to consider, and it has no solution for $\lambda<-1$. There is also no solution unless $[\sigma \pm \sqrt{1+\lambda}]$ is positive. When this condition is satisfied, there is a solution that depends on the function $a$, a choice of which amounts to a choice of gauge. For $a=1$ we find the metric
\be
ds^2 = e^{2\alpha r}\left(-dt^2 + d\theta^2\right) + dr^2\,,
\ee
which is just an adS vacuum (in the absence of any identifications). Thus all `domain-wall' solutions of CGMG (of the specified type) are actually locally equivalent to an adS vacuum.

\section{Black holes}  \setcounter{equation}{0}

We may fix the gauge invariance of the energy functional
(\ref{redfull}) by setting \be \rho(r)=r\, . \ee From (\ref{ward})
we see that the $\rho$ equation of motion is implied by the other
equations of motion when $\rho=r$, so that this is a permissible
gauge choice.  The gauge-fixed energy functional is \be\label{NMG2}
E =  \int \! dr \left\{ 2\lambda m^2 br - \sigma ab' + \frac{1}{2m^2
r} \left[ \left(ba' + 2ab'\right)  \left( rc' + a'\right)  +bc^2 +
2\left(ab\right)' c\right] \right\}\, . \ee Note that this involves
only the variables and their first derivatives, which was made
possible by the introduction of the `auxiliary' variable $c$.  The
equation of motion for $c$ is \be\label{ceq} c = \frac{r}{2b}
\left(ba' + 2ab'\right)' - \frac{\left(ab\right)'}{b}\, . \ee We can
use this to simplify the $a$ equation of motion  to
\be\label{simpleeq} r\left(2m^2 \sigma b' + bc'{}' - b' c' \right) +
ba'{}' - b' a' + 2bc' = 0\, . \ee Finally, the  $b$ equation of
motion is
\begin{eqnarray}\label{bee}
0 &=& - 4\lambda m^4 r^3 +
\left(-2\sigma m^2 a'  + a' c' + 2ac'{}'\right) r^2 \nonumber \\
&& +  \left[\left(a'\right)^2 + 2a\left(a'{}' + c'\right) -c^2 \right] r -
 2a\left(a' +c\right) \, .
\end{eqnarray}

We shall seek  solutions of these equations with
 \be
 b\equiv 1\, .
 \ee
This means that the metric has the form
\be\label{static2}
ds^2 = -a dt^2 + a^{-1} dr^2 + r^2 d\theta^2\, ,
\ee
and that the Einstein tensor is
\be
G = \frac{aa'}{2r} dt^2 + \frac{a'}{2ar} dr^2 + \frac{r^2 a'{}'}{2} d\theta^2\, .
\ee
It follows that all solutions with $G_{\mu\nu}=-\Lambda g_{\mu\nu}$ have
$a'=-2\Lambda r$. More generally, any solution with
\be\label{localadS}
a= a_0 -\Lambda r^2
\ee
is locally isometric to dS$_3$ if $\Lambda>0$ and to adS$_3$ if $\Lambda<0$.  Any solution not of this form will
not be locally isometric to (a)dS.

Given that $b=1$, one may show that  the equations  (\ref{ceq}) and (\ref{simpleeq}) imply that
  \be
 a= a_0 + a_1 r + \frac{1}{2} a_2 r^2 - \alpha\left(r\log r -r\right) \, , \qquad
 c= \frac{1}{2} r a'{}' - a'\, ,
 \ee
 where $(a_0,a_1,a_2)$ and $\alpha$ are constants. Substitution into (\ref{bee}) yields
 \be
 \alpha=0\, , \qquad a_1\left(a_2 -4 m^2 \sigma\right) =0\, , \qquad
 a_2 = 4m^2\left[\sigma \pm \sqrt{1+\lambda}\right]\, .
 \ee
 We see that there is no solution when $\lambda <-1$. For $\lambda>-1$ we also have $a_1=0$, so
$a$ takes  the form (\ref{localadS}) with $\Lambda= -2m^2
\left[\sigma \pm \sqrt{1+\lambda}\right]$. For $a_0=1$ we recover
the (a)dS vacua. In an adS vacuum we may choose $a_0<0$ to get a
static BTZ black hole with mass parameter $M=-a_0$. We do not expect
the energy to coincide with $M$ but we do expect the energy to be
positive when the entropy is positive. We return to this point in
section \ref{CFTsec}.

 \subsection{New black holes at $\lambda=-1$}

When  $\lambda=-1$ we do not need to set $a_1=0$, so  in this case the metric has the form (\ref{static2})  but with
\be
a = a_0 + a_1 r  + 2\sigma m^2 r^2 \, .
\ee
For $a_1\ne0$ these solutions are not locally (a)dS, and for  $a_1^2\ge 8 \sigma m^2 a_0$, we have
\be
a = 2m^2\sigma \left(r-r_+\right) \left(r- r_-\right)\, , \qquad
r_\pm = \frac{1}{4 m^2} \left[ \sigma a_1 \pm \sqrt{a_1^2  - 8\sigma m^2 a_0}\right]\, ,
\ee
for real $r_\pm$.

For $\sigma=1$ the spacetime is asymptotically adS with $\Lambda= -2m^2$. When $r_\pm$ are real and $r_+>0$ we have an asymptotically adS black hole with horizon at $r=r_+$. There will also be an `inner' horizon at $r=r_-$ if $r_->0$. As long as $r_-\ne r_+$ the black hole will have a non-zero surface gravity but the limiting case $r_-=r_+=r_0$
yields an extremal black hole with zero surface gravity. The extremal black hole metric is
\be
ds^2 = - 2m^2  \left(r-r_0\right)^2 dt^2 + \frac{dr^2}{2m^2\left(r-r_0\right)^2} + r^2d\theta^2\, .
\ee
Near the horizon at $r=r_0$ this metric becomes the metric of  adS$_2 \times S^1$.  In the following subsection, we verify directly that this near-horizon spacetime is a solution of the CGMG field equations at $\lambda=-1$ and $\sigma=1$.

For $\sigma=-1$, we need real $r_\pm$ in order to have a static
region in which $a>0$. When $r_+>0$ the surface $r=r_+$ is analogous
to the cosmological horizon of dS space. If $r_->0$ too then we have
a black hole in this `cosmological' spacetime with horizon at
$r=r_-$. As for $\sigma=1$, we can again consider the limiting case
in which $r_+=r_-=r_0$ but the static region shrinks to the now
unique horizon at $r=r_0$. In terms of the new coordinates $(\tilde
t,\tilde r)$ defined by \be e^{\sqrt{2}\, m \tilde t} = r-r_0\, ,
\qquad \tilde r = \sqrt{2}\, mt\, , \ee one finds that the metric is
\be ds^2 = - d\tilde t^2 + e^{2\sqrt{2}\, m\tilde t} d\tilde r^2 +
\left(r_0 + e^{\sqrt{2}\, m\tilde t} \right) d\theta^2\, . \ee This
is not a static metric but in the limit that $\tilde t\to -\infty$
(which is  a near-horizon limit in which $r\to r_0$), it reduces to
\be ds^2 = - d\tilde t^2 + e^{2\sqrt{2}\, m\tilde t} d\tilde r^2 +
r_0^2  d\theta^2\, . \ee This is locally isometric to the
`Kaluza-Klein' metric  dS$_2 \times S^1$. We shall  verify below
that this is a  solution of the CNMG equations for $\lambda=-1$ and
$\sigma=-1$.

\subsection{Kaluza-Klein vacua at $\lambda=-1$}

The gauge choice $\rho=r$ is inconvenient if we wish to seek solutions for which the metric has an $S^1$ factor, since these are most easily investigated by considering $\rho'=0$. We stress that this is a matter of convenience; any solution that we find with $\rho'=0$ must  correspond to a solution in the $\rho=r$  gauge, but it will be a solution with $b\ne1$ and will be expressed in different coordinates. To avoid this we return to the gauge-invariant energy
function (\ref{redfull}) and make the alternative gauge choice $b=1$. Strictly speaking, this is not an admissible gauge choice because only the derivative of the $b$ equation of motion is implied by the other equations of motion when $b=1$, as one can see from (\ref{ward}). This means that substitution of $b=1$ into (\ref{redfull}) is not permitted. We therefore substitute both $b=1$ and $\rho=1$ into the equations of motion of (\ref{redfull}). The $c$ and $a$ equations imply that
\be\label{caeq}
2c = a'{}' \, , \qquad c'{}' =0\, .
\ee
Using this to simplify the $b$ and $\rho$ equations we deduce that
\be\label{brhoeq}
\left(a'{}'\right)^2 = 3a' a'{}'{}' + 4m^2 \sigma a'{}' \, , \qquad \left(a'{}'\right)^2
= 2a' a'{}'{}' -16\lambda m^4\, .
\ee
The equations (\ref{caeq}) imply that $a$ is a cubic polynomial in $r$. Using this information in (\ref{brhoeq}) we
deduce (i) that $a$ is actually a quadratic polynomial in $r$ with $a'{}' = -4m^2\lambda\sigma$, and (ii) that
$\lambda(1+\lambda)=0$.

When $\lambda=0$, the quadratic polynomial $a$ is actually linear.
This means that the metric is flat, as  can be seen from the fact
that
\be
dx^\mu dx^\nu R_{\mu\nu} \big|_{\rho=b=1} = -\frac{1}{2}
a'{}' \left[ -a dt^2 + a^{-1} dr^2\right] \, .
\ee
So the only non-trivial case is $\lambda=-1$. We then have the metric
\be
ds^2 =
-a dt^2 + a^{-1} dr^2 + d\theta^2 \, , \qquad a= a_0 + a_1 r  +
2m^2\sigma\, r^2\, .
\ee
Its Ricci tensor is
\be
dx^\mu dx^\nu
R_{\mu\nu} = -2m^2\sigma \left[ -a dt^2 + a^{-1} dr^2\right]\, .
\ee
The metric is therefore locally isometric to adS$_2 \times S^1$ when $\sigma =1$, which is the KK solution found previously in \cite{Clement:2009gq},  and to dS$_2\times S^1$ when $\sigma=-1$.

\section{Linearization}
\setcounter{equation}{0}\label{linearization}

We will now find the field theories obtained by linearization about an arbitrary maximally-symmetric vacuum of the CNMG model defined by the action (\ref{CNMG}). We write the metric as
\be
g_{\mu\nu} = \bar g_{\mu\nu} + \kappa h_{\mu\nu}
\ee
and define
\be
h= \bar g^{\mu\nu} h_{\mu\nu} \, , \qquad h_\mu = \nabla^\nu h_{\mu\nu}\, ,
\ee
where $\nabla$ is the covariant derivative constructed with the background metric, which we
take to satisfy
\be
\bar R_{\mu\nu} \equiv R_{\mu\nu} (\bar g) = 2\Lambda \bar g_{\mu\nu}  \qquad \left(\Rightarrow \bar R\equiv  \bar g^{\mu\nu} \bar R_{\mu\nu}  =6\Lambda \right)\, .
\ee
One finds that the full Ricci tensor has the expansion
\be
  R_{\mu\nu} \ = \ 2\Lambda \bar{g}_{\mu\nu} +\kappa
  R_{\mu\nu}^{(1)}+\kappa^2 R_{\mu\nu}^{(2)}+{\cal O}(\kappa^3)\, ,
 \ee
where
 \be
  R_{\mu\nu}^{(1)} \ = \ -\frac{1}{2}\left(\nabla^2 h_{\mu\nu}- 2\nabla_{(\mu}h_{\nu)}
  + \nabla_{\mu}\nabla_{\nu}h-6\Lambda h_{\mu\nu}+2\Lambda\bar{g}_{\mu\nu}h\right)\, .
 \ee
The explicit form of $R_{\mu\nu}^{(2)}$ is not required for our analysis, but only its
trace, which is given by
 \bea\label{Rsquare}
  \bar{g}^{\mu\nu}R_{\mu\nu}^{(2)} \ = \
  \frac{1}{2}h^{\mu\nu}\left(R_{\mu\nu}^{(1)}-\tfrac{1}{2}R^{(1)}\bar{g}_{\mu\nu}\right)+\text{total derivative}\;.
 \eea
Here we denote by $R^{(1)}$ the trace of $R_{\mu\nu}^{(1)}$ in the
background metric. The Einstein tensor is given by
 \bea
  G_{\mu\nu} \ = \ -\Lambda \bar{g}_{\mu\nu}+\kappa
  G_{\mu\nu}^{(1)}+\kappa^2G_{\mu\nu}^{(2)}+{\cal O}(\kappa^3)\;,
 \eea
where
 \begin{eqnarray}\label{lineinst}
  G_{\mu\nu}^{(1)} &=&  R_{\mu\nu}^{(1)}-\tfrac{1}{2}R^{(1)}\bar{g}_{\mu\nu}
  -3\Lambda h_{\mu\nu}+\Lambda h\bar{g}_{\mu\nu}\;, \\ \nonumber
  G_{\mu\nu}^{(2)} &=&  R_{\mu\nu}^{(2)}-\ft12 R^{(1)}h_{\mu\nu}
  -\ft12(R^{(2)}-h^{\rho\sigma}R_{\rho\sigma}^{(1)})\bar{g}_{\mu\nu}-\Lambda
  h^{\rho\sigma}h_{\rho\sigma}\bar{g}_{\mu\nu}+\Lambda h
  h_{\mu\nu}\;.
 \end{eqnarray}
We also have to expand the auxiliary field $f_{\mu\nu}$. For
non-flat backgrounds, there will be a non-vanishing background value
for $f_{\mu\nu}$ determined by (\ref{fsol}). It turns out to be
convenient to expand the auxiliary tensor field $f_{\mu\nu}$ as
 \be\label{fexpand}
  f_{\mu\nu} \ = \   \frac{1}{m^2}\left\{ \Lambda\left[ \bar{g}_{\mu\nu}  + \kappa h_{\mu\nu} \right] -
 \kappa k_{\mu\nu}\right\}  + {\cal O}(\kappa^2)\, ,
 \ee
where $k_{\mu\nu}$ is an independent symmetric tensor fluctuation
field.

At this stage it is worthwhile to discuss the presence of gauge
symmetries in this analysis. The full non-linear theory is invariant
under
 \begin{eqnarray}
  \delta_{\xi}g_{\mu\nu} &=& D_{\mu}\xi_{\nu}+D_{\nu}\xi_{\mu}\;, \\
  \nonumber
  \delta_{\xi}f_{\mu\nu} &=&
  \xi^{\rho}\partial_{\rho}f_{\mu\nu}+\partial_{\mu}\xi^{\rho}f_{\nu\rho}
  +\partial_{\nu}\xi^{\rho}f_{\mu\rho}\;,
 \end{eqnarray}
where $D_{\mu}$ denotes the full covariant derivative. Next, we
expand the diffeomorphism parameter according to
 \bea
  \xi_{\mu} \ = \ \bar{\xi}_{\mu}+\kappa \epsilon_{\mu}\;.
 \eea
Since the background metric $\bar{g}_{\mu\nu}$ is non-dynamical we
are restricted to diffeomorphisms that are also background isometries, which restricts
$\bar{\xi}_{\mu}$ to be a background Killing vector field. For the metric
fluctuations one finds the standard gauge transformations
 \bea\label{lingauge}
  \delta_{\epsilon}h_{\mu\nu} \ = \
  \nabla_{\mu}\epsilon_{\nu}+\nabla_{\nu}\epsilon_{\mu}\, ,
 \eea
while the fluctuations of the auxiliary field in (\ref{fexpand})
have been defined such that $k_{\mu\nu}$ is gauge invariant. In
contrast to the expansion about flat space, the linearized Einstein
tensor is not invariant by itself but only in a combination that
appears in the linearized field equation. Specifically, expansion of
the  EH action plus cosmological term yields the action
 \bea\label{linEin}
  S  = -\frac{\sigma}{2}\int d^3 x\sqrt{\bar{g}}\; h^{\mu\nu}{\cal
  G}_{\mu\nu}(h)\;,
 \eea
where ${\cal G}$ is the self-adjoint tensor-valued linear differential operator defined by
 \bea\label{EinsteinDS}
 \begin{split}
  {\cal G}_{\mu\nu}(h) \ &\equiv \ G_{\mu\nu}^{(1)}(h)+\Lambda
  h_{\mu\nu} \\
  \ &= \ R_{\mu\nu}^{(1)}-\ft12 R^{(1)}\bar{g}_{\mu\nu}-2\Lambda
  h_{\mu\nu}+\Lambda h\bar{g}_{\mu\nu}\, .
 \end{split}
 \eea
The  invariance  of ${\cal G}(h)$ under (\ref{lingauge}) may be verified using the  relation
 \bea
  [\nabla_{\mu},\nabla_{\nu}] V_{\rho} \ = \
  \Lambda\left(\bar g_{\mu\rho}V_{\nu} - \bar g_{\nu\rho}V_{\mu}\right)\;.
 \eea
The action (\ref{linEin}) is therefore gauge-invariant despite the presence, for $\Lambda\ne0$,  of a
mass-like term quadratic in $h$.

We are now ready to linearize (\ref{CNMG}) around a
maximally-symmetric vacuum. The terms linear in $1/\kappa$ cancel as
a consequence of (\ref{Lambdarel}). The quadratic,
$\kappa$-independent, terms in the Lagrangian are
 \bea\label{quadaction}
  {\cal L}_{2} = \frac{\left(\Lambda - 2m^2\sigma\right)}{4m^2}\, h^{\mu\nu}{\cal
  G}_{\mu\nu}(h)-\frac{1}{m^2}k^{\mu\nu}{\cal G}_{\mu\nu}(h)
  -\frac{1}{4m^2}(k^{\mu\nu}k_{\mu\nu}-k^2)\, .
 \eea
 When  $\Lambda=2m^2\sigma$, which can happen only if $\lambda=3$, the first term is absent and we may use the self-adjointness of the tensor operator ${\cal G}$ to rewrite  the second term so that
 \bea\label{quadaction2}
m^2  {\cal L}_{2} =  - h^{\mu\nu}{\cal G}_{\mu\nu}(k)
  -\frac{1}{4}\left(k^{\mu\nu}k_{\mu\nu}-k^2\right)\, .
 \eea
Below we will show that this propagates massive `photons' rather than massive gravitons.

 We now proceed on the assumption that $\Lambda \ne 2m^2\sigma$. The fields $h$ and $k$ can then be decoupled by the  field redefinition,
 \bea\label{fieldred}
  \bar{h}_{\mu\nu} \ = \ h_{\mu\nu}
  - \frac{2}{\left(\Lambda - 2m^2\sigma\right)}\, k_{\mu\nu}\;,
 \eea
which yields
 \bea\label{result2}
  {\cal L}_{2} =
  \frac{\left(\Lambda - 2m^2\sigma\right)}{4m^2}\, \bar{h}^{\mu\nu}{\cal G}_{\mu\nu}(\bar{h})-\frac{1}{m^2(\Lambda -2m^2\sigma)}
  k^{\mu\nu}{\cal
  G}_{\mu\nu}(k)-\frac{1}{4m^2}\left(k^{\mu\nu}k_{\mu\nu}-k^2\right)\, .
 \eea
The first term involving $\bar{h}_{\mu\nu}$ is the Einstein-Hilbert
action linearized about the vacuum, and therefore it does not
propagate physical degrees of freedom.  The relevant part of the
quadratic Lagangian is therefore \be\label{result3}
  {\cal L}_{2}(k)  =\frac{1}{m^2\left(\Lambda -2m^2\sigma\right)} \left\{ -\frac{1}{2}
  k^{\mu\nu}{\cal G}_{\mu\nu}(k) - \frac{1}{4} M^2
  \left(k^{\mu\nu}k_{\mu\nu}-k^2\right)\right\}\;,
  \ee
  where
  \be\label{Mvalue}
  M^2 = - \sigma m^2 + \frac{1}{2}\Lambda\, .
  \ee
Here we recall that ${\cal G}_{\mu\nu}(k)$ is defined as in
(\ref{EinsteinDS}), with $h_{\mu\nu}$ replaced by $k_{\mu\nu}$.

\section{Unitarity and Stability}
\setcounter{equation}{0}

We turn now to the analysis of unitarity for the cases $\sigma=\pm
1$. We have implicitly supposed that $m^2>0$ up to now but the
linearization actually applies also to $m^2<0$ and so in the
following we inspect this case as well. Leaving aside the special
case that leads to (\ref{quadaction2}), which we discuss below, we
see from (\ref{result3}) that ghosts are avoided if
 \bea\label{ineq}
  m^2(\Lambda-2m^2\sigma) \ > \ 0 \;.
 \eea
In addition $M^2$ has to satisfy a bound in order to avoid tachyons. About flat space ($\Lambda=0$) we require $M^2\ge0$,  so both ghosts and tachyons are avoided for $\sigma=-1$ (and $m^2>0$).  From
(\ref{Lambdarel}) we see that $\Lambda=0$ is possible only when $\lambda=0$, so we have recovered the result of   \cite{Bergshoeff:2009hq} that
(`wrong-sign') NMG with $\lambda=0$ is equivalent at the linearized level, when expanded about its Minkowski vacuum,  to the (3D) Pauli-Fierz  theory for massive spin 2, and is therefore a unitary interacting and generally covariant extension of that field theory.

On dS space, not only may $M^2$ not be negative but small positive values
below a  critical value are also forbidden \cite{Higuchi:1986py}.
To be specific, one must have
 \bea\label{dSbound}
  M^2 \ \geq \ \Lambda\;.
 \eea
From (\ref{Mvalue}) one sees that this bound is equivalent to \be
\Lambda \ \le \ -2m^2 \sigma \, . \ee This cannot be satisfied  for
dS vacua with $\sigma=1$ but is satisfied for  $\sigma=-1$ if
$\lambda<0$  and if one chooses the dS vacuum with the smaller value
of $\Lambda$.  When $\lambda=-1$, the bound may be saturated; this
is another special case that will be discussed below.

For adS vacua it is well known that unitarity allows scalar fields
to have a negative mass squared, provided that the
Breitenlohner-Freedman (BF) bound is satisfied
\cite{Breitenlohner:1982bm}. For $D=3$ this bound is just
(\ref{dSbound}) \cite{Mezincescu:1984ev}. It has been argued that
the same bound applies to spin 2 (see e.g.
\cite{Carlip:2008jk,Gover:2008sw}). Although we are not aware of any
direct proof from unitarity of this spin 2 mass bound,  we will
allow negative $M^2$ in what follows in order to examine the
possible consequences.

\subsection{The special cases $\lambda=3$ and $\lambda=-1$}\label{subsec:lam3}

For $\lambda=3$ the cosmological constant can take the value
$\Lambda=2m^2\sigma$, for which the linearized action is given by
(\ref{quadaction2}). It is clear from this result that the metric
perturbation $h$ is now a Lagrange multipler for the constraint
${\cal G}(k)=0$. The solution of this constraint is
\be k_{\mu\nu} =
2\nabla_{(\mu} A_{\nu)}\;.
\ee
If we now substitute for $k$ in (\ref{quadaction2}) we arrive at the linearized Lagrangian
 \be
 {\cal L}_{2} = -\frac{1}{4} F^{\mu\nu}F_{\mu\nu} +4m^2\sigma A^{\mu}A_{\mu}\, ,
 \qquad F_{\mu\nu} \ = \ 2\partial_{[\mu}A_{\nu]}\;.
\ee
This is just the Proca Lagrangian for a massive spin 1 field $A$,
propagating the two helicities $\pm1$ with mass squared
$-8m^2\sigma$.  Unitarity evidently requires $m^2>0$, so the spin 1 modes are tachyons
for $\sigma=1$ but physical for $\sigma=-1$. Thus,  linearization about the adS vacuum of
the $\lambda=3$ and $\sigma=-1$ model yields a unitary stable theory for massive spin 1 in
the adS background.

We next discuss $\lambda=-1$, which on dS corresponds to saturating
the bound (\ref{dSbound}). This point is special in that there is an
enhancement of gauge symmetry, first discussed by Deser and
Nepomechie \cite{Deser:1983mm}. To be precise, when $M^2=\Lambda$
the three-dimensional Pauli-Fierz Lagrangian  (\ref{result3})  has
the following gauge invariance with infinitesimal  scalar parameter
$\zeta$ \cite{Tekin:2003np}:
 \bea\label{exoticgauge}
  \delta_{\zeta}k_{\mu\nu} \ = \
  \nabla_{\mu}\nabla_{\nu}\zeta+\Lambda\bar{g}_{\mu\nu}\zeta\, .
 \eea
This phenomenon is usually discussed for dS space since the
condition $M^2=\Lambda$ implies that $M^2<0$ on adS (although the
conjectured spin 2 version of the BF-type bound would still be
satisfied; in fact, saturated). For dS space, the enhancement of
gauge symmetry is related to the possibility of `partially massless'
fields \cite{Deser:2001pe,Deser:2001us}. It  implies that the number
of propagating modes is one less  than the generic massive case, and
hence that the theory propagates just one mode rather than two.
Moreover, this one mode does not have a well-defined helicity in 3D
and thus it resembles more a massless than a massive
mode.\footnote{We thank Andrew Waldron for discussions on this
point.} An interesting question with regard to the consistency of
the theory at $\lambda=-1$ is whether this additional gauge symmetry
descends from a non-linear symmetry of the full theory. We leave
this for future work.

Summarizing, for generic values of $\lambda$, CNMG propagates two
massive graviton modes about the maximally symmetric vacua, with the
exception of $\lambda=-1$ and $\lambda=3$, for which there is a
single partially massless mode and a massive vector, respectively.

\subsection{AdS vacua}
We are now ready to discuss unitarity for the various cases: Let us
first start with the adS case, i.e., with $\Lambda<0$. We
distinguish four cases depending on the signs of $\sigma$ and $m^2$.
In the cases for which $m^2<0$ we use the new mass parameter $\tilde
m$ defined by \be \tilde m^2 =- m^2\, . \ee Our  results,  both for
adS and dS,  are summarized in the figures, together with the results
for the central charges to be discussed in the following subsection.

\begin{itemize}

\item { \underline{$m^2>0$, $\sigma=-1$}}.

This choice corresponds to the original one of
\cite{Bergshoeff:2009hq}. The condition (\ref{ineq}) requires
 \bea\label{rightbound}
  \Lambda \ > \ -2m^2 \;.
 \eea
{}From the explicit relation between $\Lambda$ and $m^2$ as implied
by (\ref{Lambdarel}) it follows that $\lambda<3$.  In total we find that there are adS vacua around which
modes with helicity $\pm 2$ propagate unitarily for the following
range of the cosmological parameter in the action,
 \bea
  0 \ < \ \lambda \ < \ 3\;.
 \eea
Moreover, since the mass term is manifestly positive, there are no
tachyonic modes and so the vacua are always stable.

As we saw in  subsection \ref{subsec:lam3}, the limiting case
\be
\lambda=3
\ee
also yields a unitary theory without tachyons but the propagating massive modes
have spin 1 rather than spin 2.

\begin{figure}\label{fig1} {\small
\centering \psfrag{supercaption}{
} \psfrag{0}{0} \psfrag{1}{1} \psfrag{2}{2} \psfrag{3}{3}
\psfrag{min1}{-1} \psfrag{min2m2}{$-2m^2$} \psfrag{2m2}{$2m^2$}
\psfrag{4m2}{$4m^2$} \psfrag{bigL}{$\Lambda \uparrow$}
\psfrag{l}{$\lambda \rightarrow$}
\includegraphics[width=11truecm]{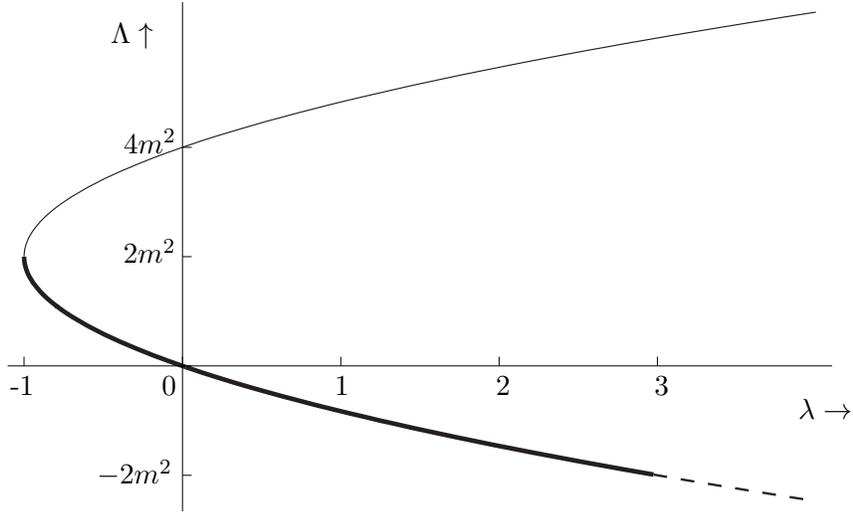}
\caption{The case $\sigma=-1,\, m^2>0$, where the fat line indicates
the unitary region, the dashed line the region with positive central
charge.} }
\end{figure}



\item {\underline{$m^2>0$, $\sigma=1$}}.

In the case of a negative value of $\Lambda$ it is not possible to
satisfy (\ref{ineq}), and therefore this theory necessarily contains
ghosts.

\item{\underline{$m^2<0$, $\sigma=1$}}.

The bound (\ref{ineq})
implies
 \bea
  2\tilde m^2+\Lambda \ < \ 0\;.
 \eea
With the explicit form of $\Lambda$ this condition amounts to $2<\mp
\sqrt{1+\lambda}$. Since the lower sign gives rise to the adS vacua,
this can be satisfied if we set
 \bea\label{lowerineq}
  \lambda \ > \ 3\;.
 \eea
We also have to check the BF-type bound as the mass parameter $m^2$
is now negative. However, by (\ref{Mvalue}) the bound becomes
$2\tilde m^2\geq \Lambda$, which is always satisfied for negative
$\Lambda$.

We should mention that the equations of motion corresponding to the
action for this choice of signs were analyzed in \cite{Liu1}.
Although the field equations appear perfectly physical for
$\lambda<3$ (even without invoking a spin 2 BF-type bound) the
required overall sign for the action implies that the propagated
modes are ghosts. This conclusion agrees with the finding of
\cite{Liu1}, which we confirm in the following section, that
gravitons have negative energy whenever the central charge is
positive.

\begin{figure}
\centering \psfrag{supercaption}{
} \psfrag{0}{0} \psfrag{1}{1} \psfrag{2}{2} \psfrag{3}{3}
\psfrag{min1}{-1} \psfrag{min2m2}{$-2m^2$} \psfrag{2m2}{$2m^2$}
\psfrag{4m2}{$4m^2$} \psfrag{min4m2}{$-4m^2$}\psfrag{bigL}{$\Lambda
\uparrow$} \psfrag{l}{$\lambda \rightarrow$}
\includegraphics[width=11truecm]{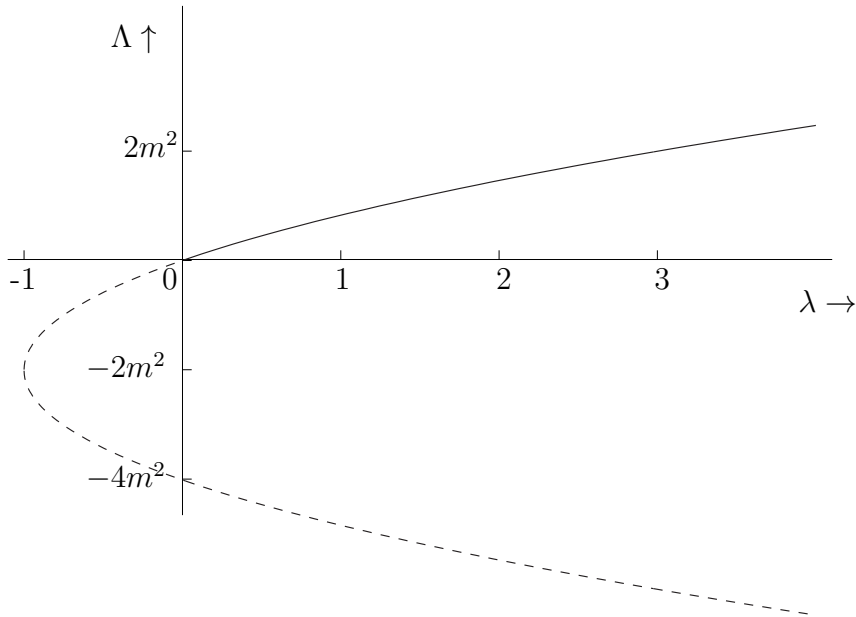}
\caption{The case $\sigma = 1,\, m^2 > 0$, with the dashed line
indicating the region with positive central charge. There is no
region with unitary gravitons.}
\end{figure}



\begin{figure}
\centering \psfrag{supercaption}{
} \psfrag{0}{0} \psfrag{1}{1} \psfrag{2}{2} \psfrag{3}{3}
\psfrag{min1}{-1} \psfrag{min2m2}{$-2{\tilde m}^2$}
\psfrag{2m2}{$2{\tilde m}^2$} \psfrag{4m2}{$4{\tilde m}^2$}
\psfrag{bigL}{$\Lambda \uparrow$} \psfrag{l}{$\lambda \rightarrow$}
\includegraphics[width=11truecm]{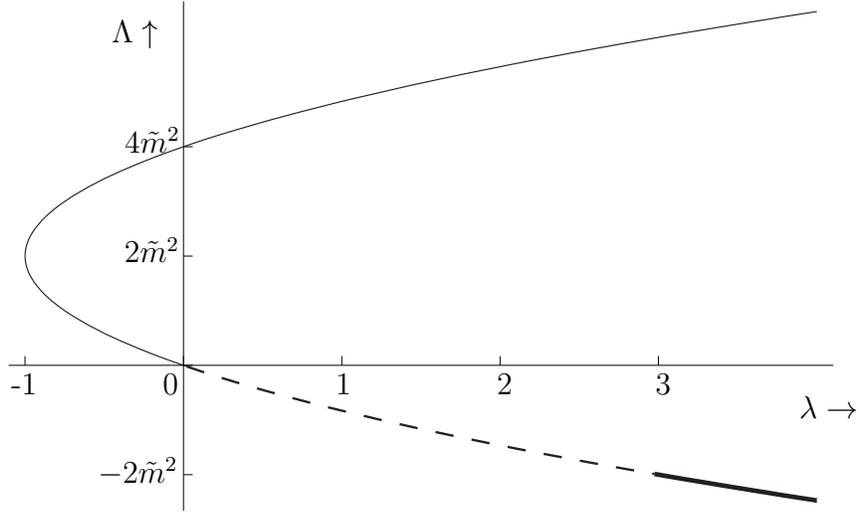}
\caption{The case $\sigma = 1,\, m^2 < 0$, with the fat line
indicating the region consistent with the BF-type bound, the dashed
line the region with positive central charge.}
\end{figure}



\item{\underline{$m^2<0$, $\sigma=-1$}}.

The condition (\ref{ineq}) amounts to $-2\tilde m^2+\Lambda<0$,
which is always satisfied. The BF-type bound, on the other hand,
reads $-2\tilde m^2\geq \Lambda$. Using the explicit form of
$\Lambda$ this means $0\geq \pm\sqrt{1+\lambda}$, which is only
satisfied for the lower sign. This corresponds to the lower branch
adS solutions and so in this region the theory is stable.

\begin{figure}
\centering \psfrag{supercaption}{
} \psfrag{0}{0} \psfrag{1}{1} \psfrag{2}{2} \psfrag{3}{3}
\psfrag{min1}{-1} \psfrag{min2m2}{$-2{\tilde m}^2$}
\psfrag{2m2}{$2{\tilde m}^2$} \psfrag{4m2}{$4{\tilde m}^2$}
\psfrag{bigL}{$\Lambda \uparrow$} \psfrag{l}{$\lambda \rightarrow$}
\psfrag{min4m2}{$-4{\tilde m}^2$}
\includegraphics[width=11truecm]{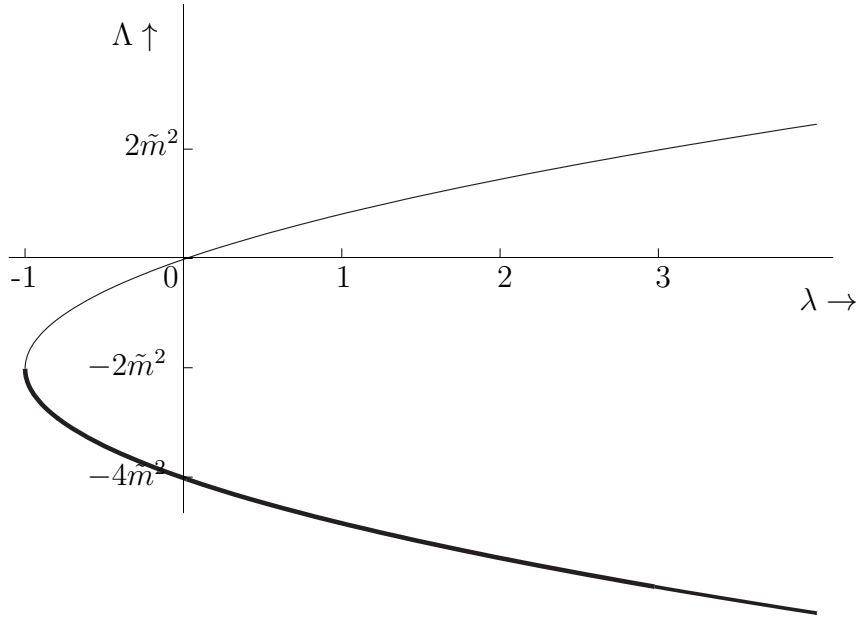}
\caption{The case $\sigma = -1,\, m^2 < 0$, where the fat line
indicates the region consistent with the BF-type bound, while there
is no region with positive central charge.}
\end{figure}



\end{itemize}

In summary, we conclude that for either sign of the Einstein-Hilbert
term there are regions of $\lambda$-space for which linearization
about an adS vacuum yields a  unitary theory,  provided both sign
choices of $m^2$ (consistent with the BF-type bound) are allowed.

\subsection{dS vacua}

Let us next analyze the de Sitter case with $\Lambda>0$. Here we
have to choose $m^2>0$ in order to avoid tachyons.

\begin{itemize}

\item \underline{$m^2>0$, $\sigma=-1$:}

The bound (\ref{dSbound}) implies together with (\ref{Mvalue})
 \bea
  m^2 \ \geq \ \frac{\Lambda}{2}\;.
 \eea
One infers that this can only be satisfied for the smaller solutions
of (\ref{Lambdarel}), which implies that precisely the lower-branch
dS solutions in the range
 \bea
  -1 \ \leq \ \lambda \ \leq \ 0\;
 \eea
are unitary. The special point $\lambda=-1$ corresponds to the gauge
symmetry enhancement, giving rise to a single `partially massless'
degree of freedom.

\item  \underline{$m^2>0$, $\sigma=1$:}

Here the bound (\ref{dSbound}) reads $-2m^2\geq \Lambda$,
which cannot be satisfied for  positive  $\Lambda$.

The point $\lambda=3$ is again special in that  the linear modes
propagating about the dS vacuum have spin 1 rather than spin 2, but
these modes are tachyonic. In other words, the dS vacuum is
unstable.

\end{itemize}
In summary, we conclude that for dS backgrounds only the original
sign choice of \cite{Bergshoeff:2009hq} leads to unitary
non-tachyonic modes, which have spin-2.

\section{Boundary CFT and central charges}\label{CFTsec}\setcounter{equation}{0}

Now we discuss some aspects of the boundary degrees of freedom
governed by a CFT and compare them with the bulk analysis of the
previous section. It has been shown by Brown and Henneaux
\cite{Brown:1986nw} that gravity theories with an $adS_3$ vacuum
generally admit an asymptotic symmetry group at the boundary which
consists of two copies of the Virasoro algebra, corresponding to the
conformal symmetry of a two-dimensional dual field theory. The
central charges of the left- and right-moving copy of the Virasoro
algebra (i.e. the holographic Weyl anomaly \cite{Henningson:1998gx})
can be expressed in terms of the 3D Newton's constant $G_3$ (related
to our parameter $\kappa$ by $\kappa^2 = 16\pi G_3$) and the adS
length $\ell$ (related to our parameter $\Lambda$ by
$\Lambda=-1/\ell^2$). In case of pure adS Einstein gravity, they are
 \bea\label{ccharge}
  c_{L} \ = \ c_{R} \ = \ c \ = \ \frac{3\ell}{2 G_3}\;.
 \eea
The central charges also encode the entropy of black holes via
Cardy's formula. Explicitly, the entropy of the BTZ black hole is
given by \cite{Kraus:2005vz}
 \bea\label{entropy}
  S \ = \ \frac{A_{\rm BTZ}}{4G_3}\Omega \;,
 \eea
where $A_{\rm BTZ}$ is the standard area of the BTZ black hole and
$\Omega=\frac{2G_3}{3\ell}c$, which is proportional to the central
charge $c$ and equal to 1 for pure adS gravity.

Next, we are going to discuss the central charges for NMG with adS
background. The systematic way to determine the central charges
would be to follow the Brown-Henneaux argument \cite{Brown:1986nw}
(see also
\cite{Henningson:1998gx,Balasubramanian:1999re,deHaro:2000xn}).
Luckily, it has been shown how for any parity-preserving
higher-derivative gravity theory admitting $adS_3$ vacua the central
charge $c$ can be derived from the general formula
\cite{Saida:1999ec,Imbimbo:1999bj,Kraus:2005vz}
 \bea\label{generalc}
  c \ = \ \frac{\ell}{2 G_3}g_{\mu\nu}\frac{\partial {\cal L}_3}{\partial
  R_{\mu\nu}}\;.
 \eea
Here ${\cal L}_3$ denotes the (higher-derivative) Lagrangian
(without the $1/\kappa^2$ pre-factor). The factors in
(\ref{generalc}) are determined such that (\ref{generalc})
reproduces the Brown-Henneaux result (\ref{ccharge}) for pure adS
gravity. Moreover, it has been shown that for a generic
higher-derivative theory admitting BTZ black hole solutions, the
entropy computed from the Wald formula coincides with the entropy
derived from the central charge (\ref{generalc}) according to
(\ref{entropy}). Applying (\ref{generalc}) to CNMG we find
 \bea\label{cchargeM}
  c \ = \ \frac{3\ell}{2 G_3}\left(\sigma-\frac{\Lambda}{2m^2}\right)
  \ = \ \frac{3\ell}{2 G_3}\left(\sigma+\frac{1}{2m^2\ell^2}\right) \;.
 \eea
In order for the boundary CFT to be unitary the central charges need
to be positive. As the central charges encode the entropy, this also
amounts to positive entropy. Here we do not attempt to compute other
physical parameters of the BTZ black hole like mass or angular
momentum from first principles, but it is reasonable to expect that
also their values will be `physical' if and only if the central
charges are positive.\footnote{See also \cite{Clement:2009gq}.} In
the following we analyze the central charges for the possible
choices of parameters:

\begin{itemize}

\item{\underline{$m^2>0$, $\sigma=-1$}.}

Here one finds that positivity of the central charge requires
$2m^2\ell^2<1$, or equivalently $\Lambda<-2m^2$. The critical value
$\Lambda=-2m^2$ is precisely the one found from the unitarity
analysis above, see (\ref{rightbound}), corresponding to the special
value $\lambda=3$, but the allowed range is the opposite. Thus,
there is no region in which both the bulk gravitons and the BTZ
black holes are well-behaved, except possibly at $\lambda=3$ where
the central charge vanishes and the bulk gravitons are replaced by
bulk `photons'.

\item \underline{$m^2>0$, $\sigma=1$}.

In this case the central charges are manifestly positive. However,
there was no region with unitarily propagating gravitons.

\item  \underline{$m^2<0$, $\sigma=1$}.

The positivity condition implies by use of the explicit expression
for $\Lambda$
 \bea
  2 \ > \ \sqrt{1+\lambda}\;,
 \eea
which is satisfied for
 \bea
  0 \ <  \ \lambda \ <  \ 3\;.
 \eea
This is again the opposite of the corresponding bound
(\ref{lowerineq}) found for unitarity in the bulk.

\item \underline{$m^2<0$, $\sigma=-1$}.

In this case the central charges are manifestly negative.

\end{itemize}
To summarize, the unitary regions of the dual CFT corresponding to
adS/BTZ backgrounds never coincide with regions in parameter space
corresponding to unitary positive-energy massive spin-2 modes in the
bulk. The extreme case is $\lambda=3$ at which the bulk modes are
given by unitary massive spin-1 excitations and where the central
charges vanish. In other words, as in TMG there is a conflict on adS
backgrounds between having `well-behaved' graviton modes or BTZ
black holes.

Potentially, this conflict might be resolved by  following the same
route as `chiral gravity' \cite{Li:2008dq}. Here it has been
conjectured for TMG with `right-sign' Einstein-Hilbert term that
there are sufficiently strong boundary conditions such that the
massive bulk modes disappear at a chiral point, where one of the
central charges vanishes; see e.g. \cite{Maloney:2009ck} for a very
recent discussion. If we consider the most general  CGMG model (with
LCS term) something similar might happen. In this case the
parity-violating Chern-Simons term leads to different left- and
right-moving sectors \cite{Kraus:2005zm}. One finds
 \begin{eqnarray}
  c_{L} =
  \frac{3\ell}{2G_3}\left(\sigma+\frac{1}{2\ell^2 m^2}+\frac{1}{\mu\ell}\right)\;,
  \qquad
  c_{R} =
  \frac{3\ell}{2G_3}\left(\sigma+\frac{1}{2\ell^2 m^2}-\frac{1}{\mu\ell}\right)\;,
 \end{eqnarray}
which allows for chiral choices with, say, $c_R=0$ and $c_L>0$. The
question is whether there are consistent boundary conditions which
indeed lead to chiral and finite charges at the boundary and which
eliminate all problematic modes in the bulk. First attempts into
this direction appeared already in \cite{Liu:2009bk}. Concerning the
special point $\lambda=3$ there might be yet another way to
establish a `chiral' theory of gravity: If the spin content is not
modified by the addition of the LCS term then there might be a choice
of parameters for which precisely one central charge vanishes while
there is no bulk graviton anyway,  but only a spin-1 excitation.

\section{Conclusions and outlook}
In this paper we have further investigated the `new massive gravity'
theory proposed in \cite{Bergshoeff:2009hq} as a generally covariant
interacting extension of the three-dimensional Pauli-Fierz theory
for massive spin 2, and to a lesser extent its generalization  to
`general massive gravity'. We have allowed for a cosmological term,
parametrized by the dimensionless constant $\lambda$ already
introduced in  \cite{Bergshoeff:2009hq}, and we have focused on new
features of the various  (a)dS vacua. With regard to both static
solutions and unitarity properties, we have found that the values
$\lambda=-1$ and $\lambda=3$ are special.

In particular, we have found a new class of asymptotically-adS black
hole solutions for $\lambda=-1$, including an extremal black hole
solution that interpolates between the adS$_3$ vacuum and an
adS$_2\times S^1$ solution. There is also an enhanced gauge
invariance of the linearized theory at  $\lambda=-1$, which implies
that one of the bulk modes can be gauged away, leaving a single
`partially massless' degree of freedom without spin.

The existence of (a)dS$_2\times S^1$ KK vacua when $\lambda=-1$
implies that the dimensional reduction of the $\lambda=-1$ models
leads to a 2D `gravitational'  theory  that has either a dS vacuum
(if $\sigma=-1$) or an adS vacuum (if $\sigma=1$). It would be of
interest to see what this 2D theory is, and to determine its
properties. In this context the recent work of \cite{Myung:2009sk}
and \cite{Kim:2009jm} may be relevant.

The models with $\lambda=3$ are also special. First, $\lambda=3$ is
the border of the unitarity region, as shown in figs.~1 and 3.
Second, at this point there are no spin-2 gravitons in the bulk but
rather massive spin-1 modes.  We also found that $\lambda=3$ is
special in our attempt to find static domain wall solutions. The
domain wall energy is a perfect square for $\lambda=3$, so its
minimization leads to a first-order Bogomol'nyi-type equation,
solutions of which yield the adS vacuum at $\lambda=3$. This
suggests that this vacuum might be supersymmetric in the
supergravity context.   It was shown in \cite{Bergshoeff:2009hq}
that  the Minkowski vacuum for $\lambda=0$ is supersymmetric, but
not enough is known about the non-linearities of the supergravity
theory to decide the issue for other vacua. The supergravity
extension is therefore still an outstanding open problem. It also
seems likely that supergravity extensions will have improved
ultra-violet behaviour; it is not unreasonable to expect an
ultra-violet  finite theory  for a sufficient number of
supersymmetries.

As in the expansion about flat space analyzed in
\cite{Bergshoeff:2009hq} we found that in order to have unitarily
propagating gravitons we have to choose the `wrong-sign'
Einstein-Hilbert term for flat space and dS vacua. Assuming a
proposed spin 2 version of the BF-type bound, we found that it was
possible to have unitarily propagating gravitons in an adS vacuum
for either sign of the EH term, but in all cases bulk unitarity
corresponds to a negative central charge for the boundary CFT, with
the possible exception of the adS vacuum of the `wrong-sign' model
with $\lambda=3$. Thus, as in cosmological TMG, there is an
incompatibility between bulk gravitons and unitarity of the boundary
CFT. As the sign of the central charge is related to the sign of the
entropy of BTZ black holes, there is an apparent  conflict between
physical bulk gravitons and BTZ black holes. We mentioned that one
resolution of the conflict might be to consider a generalization of
`chiral gravity', which would eliminate the bulk gravitons. Another
point of view might be to keep the unitary bulk gravitons but to
discard the BTZ black holes; it  has been argued for TMG that there
is a superselection argument that would exclude them
\cite{Carlip:2008jk} but for NMG this option, if it is an option,
has to be considered in the light of the fact that there are other
types of black hole.

There are clearly many further features of NMG and GMG that deserve study.
To mention just one: a connection to Ho\v{r}ava-Lifshitz gravity has recently been
proposed \cite{Cai:2009ar}.

\subsection*{Acknowledgments}
We acknowledge discussions with N.~Boulanger, S. Deser,
D.~Grumiller, M.~Henneaux, A.~Maloney, K.~Skenderis, P.~Sundell,
R.~Troncoso and A.~Waldron. We thank T.~Nutma for  making the figures.
We further thank the organizers of the
ESI workshop `Gravity in Three Dimensions' for providing a
stimulating atmosphere. This work is part of the research programme
of the `Stichting voor Fundamenteel Onderzoek der Materie (FOM)'.

\begin{appendix}
\renewcommand{\theequation}{\Alph{section}.\arabic{equation}}

\section{`K-models'} \setcounter{equation}{0}

In our linearization analysis, all results were obtained for an
arbitrary sign $\sigma$ but  they also apply if $\sigma=0$. By
setting $\sigma=0$ we effectively have a model with a cosmological
and (higher-derivative) `$K$'-term but no EH term. We shall call
these the `$K$'-models; in them, the cosmological constant is
related to our parameter $\lambda$ by \be \Lambda^2 = 4\lambda m^2\;
. \ee We will consider both signs of $m^2$. For either sign one can
choose $\lambda$ to get any desired value of $\Lambda^2\ge0$. The
choice $\lambda=0$ implies $\Lambda=0$ and hence a unique  Minkowski
vacuum. This special case yields the massless  NMG model mentioned
in the  introduction. We shall first discuss this special case in
some detail, and then comment on its `cosmological' extension to
$\Lambda^2>0$.

\subsection{$\Lambda=0$}

We noted in subsection \ref{subsec:lam3} that when $\Lambda= 2m^2\sigma$ the metric perturbation becomes
a Lagrange multiplier in the linearized theory,  imposing a constraint of vanishing linearized curvature for a
PF tensor field. This observation still applies when $\sigma=0$, in which case the metric perturbation becomes
a Lagrange multiplier in the linearized theory at  $\Lambda=0$, so  our previous analysis of NMG at $\Lambda= 2m^2\sigma$ implies that massless NMG is equivalent, for $m^2>0$,  to  a Maxwell action for a 3-vector field.
This is  on-shell equivalent to the action for a massless scalar field, so there is one
propagating mode, in agreement with \cite{Deser:2009hb}.  We should point out here that spin
is not defined for a massless particle in 3D, so one cannot really assign a spin to this one massless particle.

We will now verify these conclusions, taking as our starting point
the `pure-K' action (\ref{pureK}) in its form (\ref{masslessact})
with auxiliary PF tensor field $\tilde f$. We linearize about the
Minkowski vacuum by setting \be g_{\mu\nu} = \eta_{\mu\nu} +
h_{\mu\nu}\;, \ee where $\eta$ is the Minkowski metric and $h$ is
now a {\it dimensionless} metric perturbation. Omitting boundary
terms, we may write the resulting quadratic action as
\be\label{firstO} S_{lin}[h,f] = \frac{1}{\beta^2}\int \! d^3 x\,
\left\{ h^{\mu\nu} \left[{\cal G}(\tilde f)\right]_{\mu\nu}  -
\frac{1}{4} \left(\tilde f^{\mu\nu} \tilde f_{\mu\nu} - \tilde
f^2\right)\right\}\, , \ee where indices are now raised and lowered
with the Minkowski metric, and ${\cal G}$ is the self-adjoint
`Einstein operator' of (\ref{EinsteinDS}) for $\Lambda=0$; it can be
written as \cite{Bergshoeff:2009hq} \be {\cal
G}_{\mu\nu}{}^{\rho\sigma} = -\frac{1}{2}
\varepsilon_{(\mu}{}^{\eta\rho} \varepsilon_{\nu)}{}^{\tau\sigma}
\partial_\eta\partial_\tau \, . \ee On the one hand, elimination of
$\tilde f$ yields the quadratic approximation to the action
(\ref{pureK}). On the other hand, we may view $h$ as a Lagrange
multiplier for the constraint ${\cal G}(\tilde f)=0$, which has the
general solution \be\label{maxwellpot} {\tilde f}_{\mu\nu} =
\partial_\mu A_\nu +
\partial_\nu A_\mu \equiv 2\partial_{(\mu} A_{\nu)}\, , \ee where
$A$ is a vector field. Remarkably, the action then reduces to
\be\label{Max} S_{lin}[A] = - \frac{1}{4\beta^2}\int \! d^3 x\,
F^{\mu\nu} F_{\mu\nu}\, , \qquad F_{\mu\nu} = \partial_\mu A_\nu -
\partial_\nu A_\mu \equiv 2\partial_{[\mu} A_{\nu]}\, . \ee
This is the Maxwell action in 3D with coupling constant $\beta$, which is real when $m^2>0$.

There is a precedent for this construction \cite{Deser:1980fy}. In {\it any}
dimension, the PF mass term becomes a Maxwell Lagrangian  on solving
the constraint of  zero linearized Riemann tensor for the spin 2
field. In 3D the Riemann tensor  is zero if the Einstein tensor is
zero, so the relevant action is (\ref{firstO}).  In 4D one needs a
Lagrange multiplier that is a 4th rank tensor field with the
algebraic symmetries of the Riemann tensor, and elimination of the
PF tensor field now yields a `pure' 4th-order Lagrangian for this
4th rank tensor field. A canonical analysis of this `higher-rank
representation' of  spin 1  confirms its spin 1 content
\cite{Deser:1980fy}. The linearized `pure-K' theory is just the 3D
analog of this 4D model. One significant difference is that it  was
not clear in the 4D case how non-trivial interactions could be
introduced, whereas in 3D the fully non-linear `pure-K' theory was
our starting point. However, there are potential difficulties with
interactions even in 3D.

As stressed in \cite{Deser:2009hb}, the
quadratic approximation to $K$ is invariant under a linearized `Weyl
invariance'. In this context we note the following convenient
rewriting of the (non-linear) $K$ term,
 \bea
  S \ = \ \int d^3x\;\sqrt{|g|}\;K \ = \ -\int
  \varepsilon_{abc}\;e^{a}\wedge f^{b}\wedge f^{c}\;,
 \eea
where $f^{a}=dx^{\mu}(R_{\mu}{}^{a}-\tfrac{1}{4}R e_{\mu}{}^{a})$ is
the `conformal boost' gauge field expressed in terms of the
curvature. In the first-order form (\ref{firstO}) the Weyl
invariance is given by
 \bea\label{Weylgauge}
  \delta_{\zeta} h_{\mu\nu} \ = \ 2\zeta \eta_{\mu\nu}\;, \qquad
  \delta_{\zeta} \tilde{f}_{\mu\nu} \ = \
  -2\partial_{\mu}\partial_{\nu}\zeta\;.
 \eea
This linearized Weyl symmetry is a consequence of  the conformal
covariance property of $K$ noted in \cite{Bergshoeff:2009hq}
($\delta K \propto K$), which also implies that this `extra' gauge
invariance of the quadratic approximation is an artefact of this
approximation; it does not extend to the interacting theory. In
fact, the linearized gauge-invariance (\ref{Weylgauge}) just amounts
to the Maxwell gauge invariance of the potential introduced in
(\ref{maxwellpot}). This gauge invariance does not survive the
introduction of interactions, so there is a constraint in the full
theory that is missed on linearization. The implications of this
constraint  are not clear to us at present, so the consistency of
massless NMG is still in doubt.  We note, however, that  the
situation in this respect is much better for the  $\lambda=3$ CNMG
model since in that case there is no gauge symmetry at the quadratic
level that  could be broken by interactions.

\subsection{$\Lambda\neq 0$}
We now turn to the cosmological `K-model'  with $\lambda>0$. We take
as our starting point the model obtained from (\ref{CNMG}) by
setting $\sigma=0$. This has one dS vacuum and one adS vacuum, both
with the same (arbitrary) value of $|\Lambda|$. The linearization
analysis of sec.~\ref{linearization} can be taken over to this case
by setting $\sigma=0$ in (\ref{result3}). We observe that even
though the original action does not have an Einstein-Hilbert term, a
linearized Einstein-Hilbert term is generated in the linearization
about non-flat backgrounds. Consequently, the linearized excitations
are actually massive spin-2 modes, in contrast to the $\lambda=0$
model.

Let us now examine unitarity of the linearized field theory on
(a)dS. From (\ref{ineq}) we infer that unitarity  requires
$m^2\Lambda>0$. In the dS vacuum this means that we must have
$m^2>0$, but since $M^2 =  \Lambda/2$ the  unitarity  bound
(\ref{dSbound}) is violated. In the adS vacuum  unitarity requires
$m^2<0$. The graviton modes have $M^2= \Lambda/2 <0$ but this
satisfies the spin 2 BF-type bound, so the bulk theory appears
physical at the linearized level. However, the central charge
(\ref{cchargeM}) of the dual CFT is now
 \bea\label{cosmNMGpara}
  c \ = \ -\frac{3\ell\Lambda}{4G_3m^2}\, ,
 \eea
which is negative when $m^2<0$. Thus, as in case of
CNMG, we cannot simultaneously have unitary bulk gravitons and a
unitary boundary CFT.

\end{appendix}


\begin{thebibliography}{99}

\bibitem{Bergshoeff:2009hq}
  E.~A.~Bergshoeff, O.~Hohm and P.~K.~Townsend,
  ``Massive Gravity in Three Dimensions,''
  arXiv:0901.1766 [hep-th], to appear in Phys.~Rev.~Lett.

\bibitem{Stelle:1976gc}
  K.~S.~Stelle,
  ``Renormalization Of Higher Derivative Quantum Gravity,''
  Phys.\ Rev.\  D {\bf 16} (1977) 953.

\bibitem{Nakasone:2009bn}
  M.~Nakasone and I.~Oda,
  ``On Unitarity of Massive Gravity in Three Dimensions,''
  arXiv:0902.3531 [hep-th].

\bibitem{Deser:2009hb}
  S.~Deser,
  ``Ghost-free, finite, fourth order D=3 (alas) gravity,''
  arXiv:0904.4473 [hep-th].

\bibitem{Oda:2009ys}
  I.~Oda, ``Renormalizability of Massive Gravity in Three Dimensions,''
  arXiv:0904.2833 [hep-th].



\bibitem{Deser:1981wh}
  S.~Deser, R.~Jackiw and S.~Templeton,
  ``Topologically massive gauge theories'',
  Annals Phys.\  {\bf 140} (1982) 372
  [Erratum-ibid.\  {\bf 185} (1988\ APNYA,281,409-449.2000) 406.1988\ APNYA,281,409].

\bibitem{Deser:2002iw}
  S.~Deser and B.~Tekin,
  ``Massive, topologically massive, models,''
  Class.\ Quant.\ Grav.\  {\bf 19} (2002) L97
  [arXiv:hep-th/0203273].

\bibitem{Li:2008dq}
  W.~Li, W.~Song and A.~Strominger,
  ``Chiral Gravity in Three Dimensions'',
  JHEP {\bf 0804} (2008) 082
  [arXiv:0801.4566 [hep-th]].



\bibitem{Carlip:2008jk}
  S.~Carlip, S.~Deser, A.~Waldron and D.~K.~Wise,
  ``Cosmological Topologically Massive Gravitons and Photons,''
   Class.\ Quant.\ Grav.\  {\bf 26} (2009) 075008,
  arXiv:0803.3998 [hep-th];
  ``Topologically Massive AdS Gravity'',
  Phys.\ Lett.\  B {\bf 666} (2008) 272
  [arXiv:0807.0486 [hep-th]].



\bibitem{Kaku:1978nz}
  M.~Kaku, P.~K.~Townsend and P.~van Nieuwenhuizen,
  ``Properties Of Conformal Supergravity,''
  Phys.\ Rev.\  D {\bf 17} (1978) 3179.

\bibitem{Uematsu:1984zy}
  T.~Uematsu,
  ``Structure Of N=1 Conformal And Poincare Supergravity In (1+1)-Dimensions
  And (2+1)-Dimensions,''
  Z.\ Phys.\  C {\bf 29} (1985) 143.


\bibitem{Gover:2008sw}
  A.~R.~Gover, A.~Shaukat and A.~Waldron,
  ``Tractors, Mass and Weyl Invariance,''
  Nucl.\ Phys.\  B {\bf 812} (2009) 424
  [arXiv:0810.2867 [hep-th]].

\bibitem{Boulware:1985wk}
  D.~G.~Boulware and S.~Deser,
  ``String Generated Gravity Models,''
  Phys.\ Rev.\ Lett.\  {\bf 55} (1985) 2656.

\bibitem{Banados:1992wn}
  M.~Banados, C.~Teitelboim and J.~Zanelli,
  ``The Black hole in three-dimensional space-time'',
  Phys.\ Rev.\ Lett.\  {\bf 69} (1992) 1849
  [arXiv:hep-th/9204099].

\bibitem{Clement:2009gq}
  G.~Clement,
  ``Warped $AdS_3$ black holes in new massive gravity,''
  Class.\ Quant.\ Grav.\  {\bf 26} (2009) 105015,
  arXiv:0902.4634 [hep-th].

\bibitem{Clement:2009ka}
  G.~Clement,
  ``Black holes with a null Killing vector in new massive gravity in three
  dimensions,''
  arXiv:0905.0553 [hep-th].

\bibitem{AyonBeato:2009yq}
  E.~Ayon-Beato, G.~Giribet and M.~Hassaine,
  ``Bending AdS Waves with New Massive Gravity,''
  arXiv:0904.0668 [hep-th].

\bibitem{Oliva:2009ip}
  J.~Oliva, D.~Tempo and R.~Troncoso,
  ``Three-dimensional black holes, gravitational solitons, kinks and wormholes
  for BHT masive gravity,''
  arXiv:0905.1545 [hep-th].




  \bibitem{PSC}
  R.S. Palais, ``The principle of symmetric criticality'',
  Commun.  Math. Phys. {\bf 69}, 19 (1979)

\bibitem{Fels:2001rv}
  M.~E.~Fels and C.~G.~Torre,
  ``The principle of symmetric criticality in general relativity,''
  Class.\ Quant.\ Grav.\  {\bf 19} (2002) 641
  [arXiv:gr-qc/0108033].

\bibitem{Deser:2003up}
  S.~Deser and B.~Tekin,
  ``Shortcuts to high symmetry solutions in gravitational theories,''
  Class.\ Quant.\ Grav.\  {\bf 20} (2003) 4877
  [arXiv:gr-qc/0306114].

\bibitem{Higuchi:1986py}
  A.~Higuchi,
 ``Forbidden Mass Range For Spin-2 Field Theory In De Sitter Space-Time,''
  Nucl.\ Phys.\  B {\bf 282} (1987) 397.

\bibitem{Deser:1983mm}
  S.~Deser and R.~I.~Nepomechie,
  ``Gauge Invariance Versus Masslessness In De Sitter Space,''
  Annals Phys.\  {\bf 154} (1984) 396.

\bibitem{Tekin:2003np}
  B.~Tekin,
  ``Partially massless spin-2 fields in string generated models,''
  arXiv:hep-th/0306178.

\bibitem{Deser:2001pe}
  S.~Deser and A.~Waldron,
  ``Gauge invariances and phases of massive higher spins in (A)dS,''
  Phys.\ Rev.\ Lett.\  {\bf 87} (2001) 031601
  [arXiv:hep-th/0102166].

\bibitem{Deser:2001us}
  S.~Deser and A.~Waldron,
  ``Partial masslessness of higher spins in (A)dS,''
  Nucl.\ Phys.\  B {\bf 607} (2001) 577
  [arXiv:hep-th/0103198].

\bibitem{Breitenlohner:1982bm}
  P.~Breitenlohner and D.~Z.~Freedman,
  ``Positive Energy In Anti-De Sitter Backgrounds And Gauged Extended
  Supergravity,''
  Phys.\ Lett.\  B {\bf 115} (1982) 197.

\bibitem{Mezincescu:1984ev}
  L.~Mezincescu and P.~K.~Townsend,
  ``Stability At A Local Maximum In Higher Dimensional Anti-De Sitter Space And
  Applications To Supergravity,''
  Annals Phys.\  {\bf 160} (1985) 406.


\bibitem{Liu1}
  Y.~Liu and Y.~Sun,
  ``Note on New Massive Gravity in $AdS_3$,''
  JHEP {\bf 0904} (2009) 106,
  arXiv:0903.0536 [hep-th];


\bibitem{Brown:1986nw}
  J.~D.~Brown and M.~Henneaux,
  ``Central Charges in the Canonical Realization of Asymptotic Symmetries: An
  Example from Three-Dimensional Gravity,''
  Commun.\ Math.\ Phys.\  {\bf 104} (1986) 207.

\bibitem{Henningson:1998gx}
  M.~Henningson and K.~Skenderis,
  ``The holographic Weyl anomaly,''
  JHEP {\bf 9807} (1998) 023
  [arXiv:hep-th/9806087].

\bibitem{Balasubramanian:1999re}
  V.~Balasubramanian and P.~Kraus,
  ``A stress tensor for anti-de Sitter gravity,''
  Commun.\ Math.\ Phys.\  {\bf 208} (1999) 413
  [arXiv:hep-th/9902121].

\bibitem{deHaro:2000xn}
  S.~de Haro, S.~N.~Solodukhin and K.~Skenderis,
  ``Holographic reconstruction of spacetime and renormalization in the  AdS/CFT
  correspondence,''
  Commun.\ Math.\ Phys.\  {\bf 217} (2001) 595
  [arXiv:hep-th/0002230].

\bibitem{Saida:1999ec}
  H.~Saida and J.~Soda,
  ``Statistical entropy of BTZ black hole in higher curvature gravity,''
  Phys.\ Lett.\  B {\bf 471} (2000) 358
  [arXiv:gr-qc/9909061].

\bibitem{Imbimbo:1999bj}
  C.~Imbimbo, A.~Schwimmer, S.~Theisen and S.~Yankielowicz,
  ``Diffeomorphisms and holographic anomalies,''
  Class.\ Quant.\ Grav.\  {\bf 17} (2000) 1129
  [arXiv:hep-th/9910267].

\bibitem{Kraus:2005vz}
  P.~Kraus and F.~Larsen,
  ``Microscopic Black Hole Entropy in Theories with Higher Derivatives,''
  JHEP {\bf 0509} (2005) 034
  [arXiv:hep-th/0506176].

\bibitem{Maloney:2009ck}
  A.~Maloney, W.~Song and A.~Strominger,
  ``Chiral Gravity, Log Gravity and Extremal CFT,''
  arXiv:0903.4573 [hep-th].

\bibitem{Kraus:2005zm}
  P.~Kraus and F.~Larsen,
  ``Holographic gravitational anomalies,''
  JHEP {\bf 0601} (2006) 022
  [arXiv:hep-th/0508218].


\bibitem{Liu:2009bk}
  Y.~Liu and Y.~Sun,
 ``Consistent Boundary Conditions for New Massive Gravity in $AdS_3$,''
  arXiv:0903.2933 [hep-th];
  ``On the Generalized Massive Gravity in $AdS_3$,''
  arXiv:0904.0403 [hep-th].

\bibitem{Myung:2009sk}
  Y.~S.~Myung, Y.~W.~Kim and Y.~J.~Park,
  ``Topologically massive gravity on AdS$_2$ spacetimes,''
  arXiv:0901.2141 [hep-th].

\bibitem{Kim:2009jm}
  W.~Kim and E.~J.~Son,
  ``Central Charges in 2d Reduced Cosmological Massive Gravity,''
  arXiv:0904.4538 [hep-th].


\bibitem{Cai:2009ar}
  R.~G.~Cai, Y.~Liu and Y.~W.~Sun,
  ``On the z=4 Horava-Lifshitz Gravity,''
  arXiv:0904.4104 [hep-th].

\bibitem{Deser:1980fy}
  S.~Deser, P.~K.~Townsend and W.~Siegel,
  ``Higher Rank Representations Of Lower Spin,''
  Nucl.\ Phys.\  B {\bf 184}, 333 (1981).





\end{thebibliography}
\end{document}